\documentclass[a4paper,12pt, epsfig]{article}
\pdfoutput=1
\usepackage{epsfig}
\usepackage{epstopdf}
\usepackage{graphicx}
\usepackage{ifthen}
\usepackage{feynmp}
\usepackage{amsmath}
\usepackage[psamsfonts]{amssymb}
\usepackage{euscript}

\usepackage{latexsym}
\usepackage[arrow,matrix,curve]{xy}

\newskip\humongous \humongous=0pt plus 1000pt minus 1000pt

\newif\ifdtup

\def\,{\hspace{-.1cm}}
\def\hsp{,\hspace{.7cm}}

\def\fc#1#2 {\frac{n}{q}#1\frac{n}{q}#2}

\def\bp{\mathbf{P}}

\def\Os{|\Omega\rangle}
\def\mo#1{\int\frac{d{#1}}{2\pi}}

\renewcommand{\sin}{\textrm{sin}}

\def\exp#1{\hbox{\rm exp}\left(#1\right)}

\renewcommand{\theequation}{\arabic{section}.\arabic{equation}}
\renewcommand{\(}{\begin{equation}}
\renewcommand{\)}{end{equation} \vspace{-.05in}\linebreak}

\newcounter{saveeqn}
\newcounter{savealpheqn}

\newcommand{\alpheqn}{\setcounter{saveeqn}{\value{equation}}%
  \stepcounter{saveeqn}\setcounter{equation}{0}%
  \renewcommand{\theequation}{\mbox{\arabic{section}.\arabic{saveeqn}
\alph{equation}}}
  \renewcommand{\)}{\end{equation}}}
\def\part#1{\frac{\partial}{\partial{#1}}}%
\def\group#1{\refstepcounter{equation}\setcounter{saveeqn}
 {\value{equation}}%
  \label{#1}\setcounter{equation}{0}%
\renewcommand{\theequation}{\mbox{\arabic{section}.\arabic{saveeqn}
\alph{equation}}}
  \renewcommand{\)}{\end{equation}}}
\newcommand{\reseteqn}{\setcounter{equation}{\value{saveeqn}}%
  \renewcommand{\theequation}{\arabic{section}.\arabic{equation}}%
  \renewcommand{\)}{\end{equation}}}

\newcommand{\aalpheqn}{\setcounter{saveeqn}{\value{equation}}%
  \stepcounter{saveeqn}\setcounter{equation}{0}%
  \renewcommand{\theequation}{\mbox{
        \Alph{subsection}.\arabic{saveeqn}\alph{equation}}}
   \renewcommand{\)}{\end{equation}}}
\newcommand{\areseteqn}{\setcounter{equation}{\value{saveeqn}}%
  \renewcommand{\theequation}{\Alph{subsection}.\arabic{equation}}%
  \renewcommand{\)}{\end{equation}}}

\renewcommand{\thefootnote}{\alph{footnote}}
\renewcommand{\(}{\begin{equation}}
\renewcommand{\)}{\end{equation}}
\newcommand{\ba}{\begin{eqnarray}}
\newcommand{\ea}{\end{eqnarray}}
\newcommand{\cbp}{\mathop{\vtop{\ialign{##\crcr
   $\hfil\displaystyle{}\hfil$\crcr\noalign{\kern-13pt\nointerlineskip}
   \BIG{)}\hskip0pt\crcr\noalign{\kern3pt}}}}}
\newcommand{\pa}{\mathop{\vtop{\ialign{##\crcr
    
$\hfil\displaystyle{\oplus}\hfil$\crcr\noalign{\kern+1pt\nointerlineskip 
}
   \hspace{.08in}$^{\alpha=0}$\hskip6pt\crcr\noalign{\kern3pt}}}}}
\renewcommand{\hsp}{,\hspace{.3in}}

\catcode`\@=11
\def\vereq#1#2{\lower3pt\vbox{\baselineskip1.5pt \lineskip1.5pt
\ialign{$\m@th#1\hfill##\hfil$\crcr#2\crcr\sim\crcr}}}
\catcode`\@=12

\renewcommand{\(}{\begin{equation}}
\renewcommand{\)}{\end{equation}}

\newcommand{\beas}{\begin{eqnarray*}}
\newcommand{\eeas}{\end{eqnarray*}}

\newcommand{\bquo}{\begin{quote}}
\newcommand{\enqu}{\end{quote}}

\def\bp{{\bf{p}}}
\def\bq{{\bf{q}}}
\def\bk{{\bf{k}}}
\newcommand{\sr}{\sigma_{\rm{rel}}}

\newcommand{\beq}{\begin{equation}}
\newcommand{\eeq}{\end{equation}}
\newcommand{\bea}{\begin{eqnarray}}
\newcommand{\eea}{\end{eqnarray}}

\newskip\humongous \humongous=0pt plus 1000pt minus 1000pt

\newif\ifdtup

\catcode`\@=11

\@addtoreset{equation}{section}

\def\@normalsize{\@setsize\normalsize{15pt}\xiipt\@xiipt
\abovedisplayskip 14pt plus3pt minus3pt%
\belowdisplayskip \abovedisplayskip
\abovedisplayshortskip \z@ plus3pt%
\belowdisplayshortskip 7pt plus3.5pt minus0pt}

\def\small{\@setsize\small{13.6pt}\xipt\@xipt
\abovedisplayskip 13pt plus3pt minus3pt%
\belowdisplayskip \abovedisplayskip
\abovedisplayshortskip \z@ plus3pt%
\belowdisplayshortskip 7pt plus3.5pt minus0pt
\def\@listi{\parsep 4.5pt plus 2pt minus 1pt
      \itemsep \parsep
      \topsep 9pt plus 3pt minus 3pt}}

\relax

\catcode`@=12

\catcode`\@=11

\def\section{\@startsection{section}{1}{\z@}{3.5ex plus 1ex minus  .2ex}{2.3ex plus .2ex}{\large\bf}}

\def\thesection{\arabic{section}}
\def\thesubsection{\arabic{section}.\arabic{subsection}}

\def\appendix{\setcounter{section}{0}
 \def\thesection{Appendix \Alph{section}}
 \def\thesubsection{\Alph{section}.\arabic{subsection}}
 \def\theequation{\Alph{section}.\arabic{equation}}}
\renewcommand{\theequation}{\arabic{section}.\arabic{equation}}

\begin{document}
\def\thefootnote{\fnsymbol{footnote}}
\def\thetitle{Wave Packets Losing Their Covariance}
\def\autone{Jarah Evslin}
\def\affa{Institute of Modern Physics, NanChangLu 509, Lanzhou 730000, China}
\def\affb{University of the Chinese Academy of Sciences, YuQuanLu 19A, Beijing 100049, China}

\begin{center}
{\large {\bf \thetitle}}

\bigskip

\bigskip

{\large \noindent   Hosam Mohammed{${}^{1,2}$}{\footnote{hosam@impcas.ac.cn}},Emilio Ciuffoli{${}^{1}$}{\footnote{emilio@impcas.ac.cn}} and \autone{${}^{1,2}$}\footnote{jarah@impcas.ac.cn}}

\vskip.7cm

1) \affa\\
2) \affb\\

\end{center}

\begin{abstract}
\noindent
In neutrino physics, it is sometimes assumed that all wave packets must transform covariantly as Lorentz vectors.  We show in a simple example that even if the initial conditions of a wave packet are covariant, then evolution in a relativistic interacting theory followed by a measurement of entangled particles can lead to a wave packet which is no longer covariant.


\end{abstract}

%
\setcounter{footnote}{0}
\renewcommand{\thefootnote}{\arabic{footnote}}

\section{Introduction}

One of the main theoretical difficulties in neutrino physics is that one does not know the shape of the initial wave packets for the involved particles.  It is common to use Gaussian wave packets.  However in Ref.~\cite{naumov1} the authors introduced covariant wave packets, defined below.  In Ref.~\cite{naumov2} the authors implicitly claim that wave packets must be covariant, at least for relativistic systems.  This assumption was then included by the Daya Bay collaboration in its analysis of whether it has observed decoherence \cite{dayadec}.  This analysis is quite important as decoherence could in principle reduce the neutrino oscillation signal\footnote{This is shown in the Appendix.}, thus explaining the fact that Daya Bay observes a lower value of the mixing angle $\theta_{13}$ than most other experiments.   Without the low value of $\theta_{13}$ observed by Daya Bay, the evidence for leptonic CP-violation reported by T2K \cite{t2k} would be weakened considerably.  In addition, if decoherence was already observed by Daya Bay, JUNO's sensitivity to the neutrino mass hierarchy would be severely reduced \cite{steven}.  Therefore it is of interest to know whether wave packets really are covariant in the sense of Ref.~\cite{naumov1}.

In Sec.~\ref{boostsez} we review the transformation of wave packets under boosts, reminding the reader that these are well-defined even if the wave packet is not itself covariant.  In Sec.~\ref{modsez} we consider a simple, interacting relativistic quantum field theory and we show that even if a particle begins in a covariant wave packet, its daughters will not inhabit covariant wave packets.  In Sec.~\ref{dayasez} we comment on the Daya Bay analysis of decoherence.  Finally, applications to neutrino physics are noted in Sec.~\ref{consez}.  In the Appendix we show that decoherence would affect the value of $\theta_{13}$ obtained from a standard analysis of reactor neutrino data.

\section{Boosting a Wave Packet} \label{boostsez}

Let $|0\rangle$ be a Lorentz-invariant state in a quantum field theory in $d+1$ dimensions.  Let  $a^\dag_\bp$ be the Schrodinger picture creation operators of a scalar field $\phi$ with the usual Heisenberg algebra normalization.
Here $\bp$ is a $d$-vector, the last $d$ components of a Lorentz $(d+1)$-vector $p$ which transforms covariantly under the mass $M\neq 0$ representation of the Lorentz group and squares to $M^2$.
Then the zeroth component of $p$ is
\beq
E_\bp=\sqrt{M^2+\bp^2}.
\eeq
If the scalar field is noninteracting then the state
\beq
|\bp\rangle=\sqrt{2E_\bp}a^\dag_\bp|0\rangle \label{eaeq}
\eeq
also transforms as a Lorentz vector
\beq
U(\Lambda)|\bp\rangle=|\Lambda \bp\rangle \label{l1eq}
\eeq
where $U(\Lambda)$ is the operator on the Hilbert space which represents the Lorentz transformation $\Lambda$ and the notation $\Lambda\bp$ is shorthand for the $d$ spatial components of $\Lambda p$.  In the case of an interacting theory, there will be corrections to (\ref{l1eq}) proportional to the commutator of the interaction terms $H_I$ in the Hamiltonian with $a^\dag$.  As such corrections are subleading in $H_I$, we will ignore them below.

Define a family of wave packets indexed by the $d$-vector $\bp$
\beq
|\bp\rangle=\int\frac{(d\bk)^d}{(2\pi)^d2E_\bk}f(\bk,\bp)|\bk\rangle \label{fameq}
\eeq
where $f$ is a function.  Lorentz transforming this equation and dropping all interaction terms
\bea
U(\Lambda)|\bp\rangle&=&\int\frac{(d\bk)^d}{(2\pi)^d2E_\bk}f(\bk,\bp)U(\Lambda)|\bk\rangle
=\int\frac{(d\bk)^d}{(2\pi)^d2E_\bk}f(\bk,\bp)|\Lambda\bk\rangle \nonumber\\
&=&\int\frac{(d\Lambda\bk)^d}{(2\pi)^d2E_{\Lambda\bk}}f(\bk,\bp)|\Lambda\bk\rangle
=\int\frac{(d\bk)^d}{(2\pi)^d2E_\bk}f(\Lambda^{-1}\bk,\bp)|\bk\rangle. \label{traneq}
\eea

Following Ref.~\cite{naumov1} we say that $|\bp\rangle$ is a covariant wave packet if
\beq
U(\Lambda)|\bp\rangle=|\Lambda\bp\rangle.
\eeq
In this case
\beq
\int\frac{(d\bk)^d}{(2\pi)^d2E_\bk}f(\Lambda^{-1}\bk,\bp)|\bk\rangle=|\Lambda\bp\rangle
=\int\frac{(d\bk)^d}{(2\pi)^d2E_\bk}f(\bk,\Lambda\bp)|\bk\rangle
\eeq
and so the covariance condition is equivalent to
\beq
f(\Lambda^{-1}\bk,\bp)=f(\bk,\Lambda\bp) \label{covarcond}
\eeq
which implies that $f$ depends upon its arguments only via Lorentz scalars.  Whether or not the wave packet is covariant, it transforms according to (\ref{traneq}).   In particular, there is no clear inconsistency in a noncovariant wave packet, although in the noncovariant case a Lorentz transformation of a state $|\bp\rangle$ takes it out of the family of states (\ref{fameq}).  

The same remains true if we demand, as is done in Ref.~\cite{naumov1}, that the wave packets are actually two-parameter families parameterized by $\bp$ and a momentum standard deviation $\sigma$ and that in the limit $\sigma\rightarrow 0$ the functions $f(\bk,\bp)$ are proportional to $\delta^d(\bk-\bp)$.  For example, if
\beq
|\bp\rangle=\int\frac{(d\bk)^d}{(2\pi)^d2E_\bk}\exp{\frac{(\bk-\bp)^2}{2\sigma}}|\bk\rangle 
\eeq
then clearly
\beq
U(\Lambda)|\bp\rangle=\int\frac{(d\bk)^d}{(2\pi)^d2E_\bk}\exp{\frac{((\Lambda^{-1}\bk-\bp)^2}{2\sigma}}|\bk\rangle .
\eeq

\section{Losing Covariance in a Simple Interacting Model} \label{modsez}

The above review suggests that covariant wave packets are not required for the consistency of Lorentz transformations.  But perhaps Nature nonetheless chooses covariant wave packets?  We will now argue that this is unlikely by considering a simple relativistic quantum field theory and showing that even if one begins with a particle in a covariant wave packet, its daughter particles in an interacting quantum field theory will no longer be covariant.

Consider a (1+1)-dimensional model of three massive, real canonical scalar fields $\phi_H$, $\phi_L$ and $\psi$.  In the Schrodinger picture the fields may be decomposed as
\bea
\psi(x)&=&\int\frac{d\bp}{2\pi}\frac{1}{\sqrt{2\omega(\bp)}}\left(a_{-\bp}+a^\dagger_{\bp}\right)e^{-i\bp x}\hsp \omega(\bp)=\sqrt{m^2+\bp^2}\\
\phi_I(x)&=&\int\frac{d\bp}{2\pi}\frac{1}{\sqrt{2\Omega_I(\bp)}}\left(A_{I,-\bp}+A^\dagger_{I,\bp}\right)e^{-i\bp x}\hsp \Omega_I(\bp)=\sqrt{M_I^2+\bp^2}\nonumber
\eea
where the masses are $m$ and $M_H>M_L$.  Let $|\Omega\rangle$ be the ground state and define the Fock states\footnote{To simplify expressions below, our convention has changed from Eq.~(\ref{eaeq}) by a factor of $\sqrt{2E}$.}
\beq
|I,\bp\rangle=A^\dagger_{I,\bp}|\Omega\rangle\hsp
|\bq\rangle=a^\dagger_{\bq}|\Omega\rangle
\hsp
|I,\bp;\bq\rangle=A^\dagger_{I,\bp}a^\dagger_{\bq}|\Omega\rangle . \label{aeq}
\eeq

Let the Hamiltonian $H$ be the usual massive free field Hamiltonian $H_0$ plus an interaction term
\beq
H_I=\int dx \mathcal{H}_I\hsp \mathcal{H}_I(x)=\phi_H(x)\phi_L(x)\psi(x). \label{hi}
\eeq
\beq
H_0|H,\bp\rangle=E_{0}(\bp)|H,\bp\rangle\hsp
H_0|L,\bp;\bq\rangle=E_{1}(\bp,\bq)|L,\bp;\bq\rangle
\eeq
where we have defined the eigenvalues
\beq
E_{0}(\bp)=\Omega_H(\bp)\hsp
E_{1}(\bp,\bq)=\Omega_L(\bp)+\omega(\bq).
\eeq

Our initial condition will consist of a heavy source particle in a covariant Gaussian wave packet
\beq
|0\rangle=\int\frac{d\bk}{2\pi 2E_\bk}e^{(p-k)^2/(2\sigma)}\sqrt{2\Omega_H(k)}|H,\bk\rangle \label{initeq}
\eeq
where $\sigma$ is a parameter which determines the initial width of the wave packet and $p$ is an arbitrary $(1+1)$-vector.  The integral converges as we choose the $+-$ space time signature.   We will not normalize the states.

Our strategy will be as follows. We begin with one heavy particle $\phi_H$ in a covariant wave packet (\ref{initeq}) and we let the system evolve so that it will contain a light particle $\phi_L$ and a particle $\psi$.  We will be interested in the wave packet for the particle $\psi$. 

Let ${\bf{P}}_\psi$ project a state onto the Fock sector with exactly one $\psi$ particle.  Then, to linear order in $H_I$, the $1\psi$ state at time $t$ is \cite{noi1}
\bea
|t\rangle&=&{\bf{P}}_\psi e^{-iHt}|0\rangle
=\int\frac{d\bk}{2\pi 2E_\bk}e^{(p-k)^2/(2\sigma)}{\bf{P}}_\psi  \sum_{k=0}^\infty \frac{(-iHt)^k}{k!}\sqrt{2\Omega_H(k)}|H,\bk\rangle\\
&=&\int\frac{d\bk}{2\pi 2E_\bk}e^{(p-k)^2/(2\sigma)}\sum_{k=1}^\infty\sum_{j=0}^{k-1} \frac{(-it)^k}{k!}H_0^{j}H_IH_0^{k-j-1}\sqrt{2\Omega_H(k)}|H,\bk\rangle\nonumber\\
&=&\int\frac{d\bk}{2\pi 2E_\bk}e^{(p-k)^2/(2\sigma)}\nonumber\\
&&\times\mo{\bq}\left(\sum_{k=1}^\infty\frac{(-it)^k}{k!}\sum_{j=0}^{k-1}E_{0}(\bk)^{k-j-1}E_{1}(\bq,\bk-\bq)^{j}\right)\frac{\sqrt{2\Omega_H(k)}|L,\bq;\bk-\bq\rangle}{\sqrt{8\Omega_{H}(\bk)\Omega_L(q)\omega(k-q)}}\nonumber\\
&=&\frac{1}{2}\int\frac{d\bk}{2\pi 2E_\bk}e^{(p-k)^2/(2\sigma)}\mo{\bq}\left(\frac{e^{-iE_{1}(\bq,\bk-\bq)t}-e^{-iE_{0}(\bk) t}}{E_{1}(\bq,\bk-\bq)-E_{0}(\bk)}\right)\frac{|L,\bq;\bk-\bq\rangle}{\sqrt{\Omega_L(\bq)\omega(\bk-\bq)}}. \nonumber
\eea

We are interested in the wave function for $\psi$ but we have an entangled state of $\psi$ and $\phi_L$.  This problem is readily solved.  Following the usual logic of the wave packet formulation \cite{beuthe,giunti2012} we assume that interactions with the environment will measure $\phi_L$, which is equivalent to projecting it onto a definite state or more precisely onto a definite momentum distribution.  

For simplicity we will choose this momentum distribution to be a delta function $2\pi \delta(\bq-\tilde{\bq})$ centered on $\tilde{\bq}$, although this choice will not qualitatively affect our results.  Let the operator ${\bf{P}}_\phi$ be this projection.  Then
\bea
{\bf{P}}_\phi|t\rangle&=&\frac{1}{2}\int\frac{d\bk}{2\pi 2E_\bk}e^{(p-k)^2/(2\sigma)}\left(\frac{e^{-iE_{1}(\tilde{\bq},\bk-\tilde{\bq})t}-e^{-iE_{0}(\bk) t}}{E_{1}(\tilde{\bq},\bk-\tilde{\bq})-E_{0}(\bk)}\right)\frac{|L,\tilde{\bq};\bk-\tilde{\bq}\rangle}{\sqrt{\Omega_L(\tilde{\bq})\omega(\bk-\tilde{\bq})}}\\
&=&|L,\tilde{\bq}\rangle\otimes\frac{1}{2} \int\frac{d\bk}{2\pi 2E_\bk}e^{(p-k)^2/(2\sigma)}\left(\frac{e^{-iE_{1}(\tilde{\bq},\bk-\tilde{\bq})t}-e^{-iE_{0}(\bp) t}}{E_{1}(\tilde{\bq},\bk-\tilde{\bq})-E_{0}(\bk)}\right)\frac{|\bk-\tilde{\bq}\rangle}{\sqrt{\Omega_L(\tilde{\bq})\omega(\bk-\tilde{\bq})}}.\nonumber
\eea
After these projections, the state ${\bf{P}}_\phi|t\rangle$ is a simple tensor product of a $1\phi_L$ Fock state $|L,\tilde{\bq}\rangle$ with fixed momentum $\tilde{\bq}$ and the wavepacket  
\bea
|\psi\rangle&=& \frac{1}{2}\int\frac{d\bk}{2\pi 2E_\bk}e^{(p-k)^2/(2\sigma)}\left(\frac{e^{-iE_{1}(\tilde{\bq},\bk-\tilde{\bq})t}-e^{-iE_{0}(\bk) t}}{E_{1}(\tilde{\bq},\bk-\tilde{\bq})-E_{0}(\bk)}\right)\frac{|\bk-\tilde{\bq}\rangle}{\sqrt{\Omega_L(\tilde{\bq})\omega(\bk-\tilde{\bq})}}\nonumber\\
&=&\frac{1}{2} \int\frac{d\bk}{2\pi 2E_\bk}\left(\frac{e^{-iE_{1}(\tilde{\bq},\bk)t}-e^{-iE_{0}(\bk+\tilde{\bq}) t}}{E_{1}(\tilde{\bq},\bk)-E_{0}(\bk+\tilde{\bq})}\right)\frac{e^{(p-{\tilde{k}})^2/(2\sigma)}|\bk\rangle}{\sqrt{\Omega_L(\tilde{\bq})\omega(\bk)}} \label{wpfin}
\eea
where $\tilde{k}$ is the (1+1)-vector corresponding to the momentum $\tilde{\bk}=\bk+\tilde{\bq}$.

When $\sigma\rightarrow 0$, the state (\ref{wpfin}) is proportional to $\delta(\bk-(\bp-\tilde{\bq}))$ and so it is a wave packet $|\bp-\tilde{\bq}\rangle$ in the sense of Ref.~\cite{naumov1}.   It can be written in the form (\ref{fameq}) with
\bea
|\tilde{\bp}\rangle&=&\int\frac{d\bk}{(2\pi)2E_\bk}f(\bk,\tilde{\bp})|\bk\rangle\hsp \tilde{\bp}=\bp-\tilde{\bq}\label{ffin}\\
f(\bk,\tilde{\bp})&=&
\frac{1}{\sqrt{8}}\left(\frac{e^{-iE_{1}(\tilde{\bq},\bk)t}-e^{-iE_{0}(\bk+\tilde{\bq}) t}}{E_{1}(\tilde{\bq},\bk)-E_{0}(\bk+\tilde{\bq})}\right)\frac{e^{(p-{\tilde{k}})^2/(2\sigma)}}{\omega(\bk)\sqrt{\Omega_L(\tilde{\bq})}} \nonumber
\eea
where we have divided $f$ by $\sqrt{2\omega}$ with respect to (\ref{wpfin}) to correct for the difference in convention for $|\bk\rangle$ between Eqs.~(\ref{eaeq}) and (\ref{aeq}).
$f$ in Eq.~(\ref{ffin}) is a function of $\bk$ and $\tilde{\bp}$ because
\beq
(p-\tilde{k})^2=(\sqrt{(\tilde{\bp}+\tilde{\bq})^2+m^2}-\sqrt{(\bk+\tilde{\bq})^2+m^2})^2-(\tilde{\bp}-\bk)^2.
\eeq

\begin{figure} 
\begin{center}
\includegraphics[width=2.5in,height=1.7in]{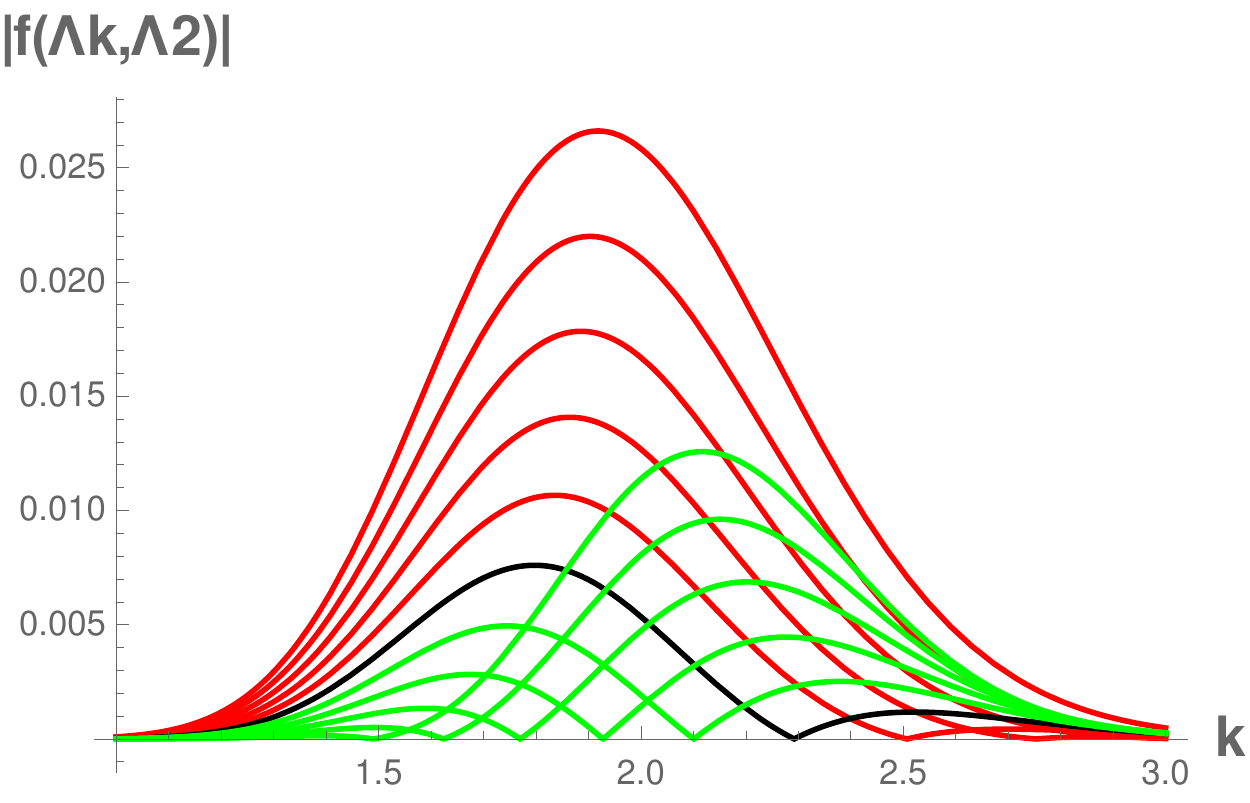}
\caption{The function $|f(\bk,\tilde{\bp})|$ at $\tilde{\bp}=2$ together with boosted wave functions $|f(\Lambda\bk,\Lambda\tilde{\bp})|$ at boost velocities from $\beta=-0.5$ to $0.5$ with steps of $0.1$, with red, black and green corresponding to $\beta<0$, $\beta=0$ and $\beta>0$ respectively.   Here we have chosen $m=M_L=\tilde{q}=1$, $M_H=10$ and $\sigma=0.01$.  We evaluated the wave function at time $t=1$.  As the curves at distinct $\beta$ are different, the wave packet is not covariant in the sense of Ref.~\cite{naumov1}.}
\label{bfig}
\end{center}
\end{figure}

The wave packet (\ref{wpfin}) is covariant only if, under an arbitrary Lorentz transformation $\Lambda$
\beq
f(\bk,\tilde{\bp})=f(\Lambda\bk,\Lambda\tilde{\bp}). \label{boost}
\eeq
In Fig.~\ref{bfig} we plot the function $f(\Lambda\bk,\Lambda\tilde{\bp})$ in Eq.~(\ref{ffin}) for different boosts of the form (\ref{boost}) and see that indeed it is not boost-invariant.  As the final wave function does not satisfy Eq.~(\ref{boost}), it is not covariant in the sense of Ref.~\cite{naumov1}.  

We thus conclude that even if Nature chooses covariant wave packets\footnote{We remind the reader that a covariant wave packet is a wave packet which, when written in the form (\ref{fameq}), satisfies (\ref{covarcond}) or equivalently (\ref{boost}).} for the initial particles,  after evolution in a relativistic quantum field theory, their daughter particles cannot be expected to have covariant wave packets.   Note that all particles began as daughter particles, and so one cannot expect initial conditions or asymptotic {\it{in}} states to be generally described by covariant wave packets.

\section{Decoherence at Daya Bay?} \label{dayasez}

What does this all have to do with the Daya Bay analysis of Ref.~\cite{dayadec}?  Daya Bay measures the $\overline{\nu}_e$ spectrum at the detectors which, given a model of the reactor fluxes, determines the oscillation probability $P$ from $\overline{\nu}_e$ to other flavors.  Daya Bay analyses fit $P$ to various models.

Our eventual goal in this project is to redo the analysis of Ref.~\cite{dayadec} with the wave packet shapes derived from the microphysics of reactor neutrino sources and scintillator detectors.  In this section we will describe why we believe that such a new analysis is warranted.  In \ref{app} we will also show how decoherence, if present but ignored, could affect the measurement of $\theta_{13}$.

Ref.~\cite{dayadec} derived the oscillation probability $P$ from a quantum mechanical treatment of neutrino wave packets.  They assumed that the neutrino is produced and detected in neutrino wave packets described by momentum space wave functions $f_P$ and $f_D$ 
\beq
f_P=\left(\frac{2\pi}{\sigma_{pP}^2}\right)^{1/4}\exp{-\frac{(p-p_P)^2}{4\sigma^2_{pP}}}\hsp
f_D=\left(\frac{2\pi}{\sigma_{pD}^2}\right)^{1/4}\exp{-\frac{(p-p_D)^2}{4\sigma^2_{pD}}}
\eeq
where $p_P$ and $p_D$ are expected momenta at production and detection and $\sigma_{pP}$ and $\sigma_{pD}$ are the corresponding expected widths.  The inverse widths are added in quadrature to yield a total momentum spread $\sigma_p$
\beq
\frac{1}{\sigma^2_p}=\frac{1}{\sigma^2_{pP}}+\frac{1}{\sigma^2_{pD}}.
\eeq
Similarly, a $\sigma$-weighted momentum $p$ is defined by
\beq
p=\frac{p_P\sigma^2_{pD}+p_D\sigma^2_{pP}}{\sigma^2_{pP}+\sigma^2_{pD}}.
\eeq
The data is used to fit a dimensionless quantity
\beq
\sr=\frac{\sigma_p}{p}.
\eeq

The oscillation probability is then defined in the usual way, by calculating a transition amplitude from the inner product of the original and final wavepacket, squaring its norm to obtain a probability and then integrating the probability over unobserved variables.  As usual, the oscillation probability depends on the momentum.  

The covariance hypothesis is inserted into the analysis via the claim that covariance implies that $\sr$ is independent of $p$.   In the rest of this section we will argue that one expects a nontrivial dependence of $\sr$ on $p$ and also that this dependence affects the final fits for $\sr$.

First, why should $\sr$ depend on $p$?  More to the point, what determines $\sr$?  The first paper on decoherence in neutrino oscillations \cite{nussinov76} already noted that the wave packet size is in part determined by pressure broadening, resulting from interactions of the source during the production process with its environment.  Reactor neutrinos of different energies result, largely, from different steps in the long decay chains of the U and Pu.  Different steps last different lengths of time and have different charges, for example, leading to different environmental interactions.  Therefore it seems extremely unlikely that distinct momenta neutrinos, which arise from distinct mixes of decays, would have identical fractional broadening.

How might this affect the analysis of Ref.~\cite{dayadec}?  The effect of $\sr$ on the oscillation probability is shown in their Fig.~1.  As expected, the effect on the near detectors, where little oscillation has occurred, is modest.  More generally, decoherence pulls the final state towards an incoherent superposition of mass eigenstates with the same fractions of each eigenstate as were present in the initial state.  In other words, it pulls the oscillation probability towards its average value.  Indeed one may observe in this figure that whenever the survival probability is above about $0.94$ the decoherence at $\sr=0.33$ pulls down the survival probability, and when it is below $0.94$ decoherence lifts up the survival probability.  

This is a general feature of decoherence and is independent of the assumptions in the paper.  However the functional form of the wave packets, and perhaps more critically the $p$-independence of $\sr$, determine just how strong this effect is at each $p$.  If for example $\sr$ is negligibly small at neutrino energies above about 4 MeV, but has a value of $\sr=0.33$ below 4 MeV, then at high energies the oscillation probability will be unchanged but at low energies the survival probability would increase.  This, if fit assuming no decoherence, would imply that the oscillation maximum occurred at a higher energy.  As a result one would find an artificially high mass splitting.  In addition, the higher value of the minimum value of the survival probability would reduce the fit value of the mixing angle $\theta_{13}$.  For example, one sees in Fig.~1 of \cite{dayadec} that $\sr=0.33$ increases the minimum value of the survival probability from about $0.918$ to $0.928$ suggesting a reduction in $\sin^2(2\theta_{13})$ of about 10\%.

There is absolutely no statistical significant mismatch between Daya Bay and accelerator neutrino measurements of the mass splitting and the mixing angle.  However, one may note that in Ref.~\cite{daya18} Daya Bay reported $\sin^2(2\theta_{13})=0.0856\pm 0.0029$ and, with the normal hierarchy, $\Delta M^2_{32}=\left(2.471^{+0.068}_{-0.070}\right)\times 10^{-3}$\ eV${}^2$ as can be compared with $\sin^2(2\theta_{13})=0.103^{+0.020}_{-0.016}$ and\footnote{The T2K determination of $\Delta M^2_{32}$ used Daya Bay's value of $\theta_{13}$.} $\Delta M^2_{32}=\left(2.343^{+0.073}_{-0.072}\right)\times 10^{-3}$\ eV${}^2$ for T2K \cite{t2k}.  Thus the best fit values are certainly consistent with the above scenario in which $\sr(p)=0$ when $p>$4 MeV but the value of $\sr(p)$ below 4 MeV lies well outside the exclusion limit $\sr<0.23$ reported in Ref.~\cite{dayadec}.

\section{Remarks} \label{consez}

Although we considered a simple model of scalar fields which enjoy a two-body decay, we believe that it is self-evident that our conclusion would also hold for more complicated models.  For example, a similar calculation could be applied to the three-body decays involving fermions which yield neutrinos.   If the initial meson or nucleus is in a covariant wavepacket, the results above then indicate that the neutrino wave packet will not be covariant.  Similarly, these initial particles were themselves created from other particles and the above calcuation may be mirrored for that process,  suggesting that the initial particles already were not described by a covariant wave packet.


\section* {Acknowledgement}

\noindent
JE is supported by the CAS Key Research Program of Frontier Sciences grant QYZDY-SSW-SLH006 and the NSFC MianShang grants 11875296 and 11675223.  EC is supported by NSFC Grant No. 11605247, and by the Chinese Academy of Sciences Presidents International Fellowship Initiative Grant No. 2015PM063.  JE and EC also thank the Recruitment Program of High-end Foreign Experts for support.

\appendix
\section{Decoherence and $\theta_{13}$}
\label{app}
Here we will show how the decoherence, if not taken into account, could affect the measurement of $\theta_{13}$ at a reactor neutrino experiment. As Daya Bay has the longest baseline of any reactor neutino experiment which measures $\theta_{13}$, one expects that this effect will be the most pronounced at Daya Bay.

We used a set-up similar to the Daya Bay experiment, considering eight 20 ton detectors, using the baselines, reactor powers, efficiencies and DAQ live times reported in \cite{An:2016ses}, Tables I and VI. Once these parameters are fixed, the total flux normalization was determined by requiring that the number of $\bar{\nu}_e$ events at the near sites (averaged over the four detectors, considering their respective livetimes and efficiencies) was equal to the one that can be obtained from Table VI of Ref. \cite{An:2016ses}. No background was taken into account. We used the Asimov data set to simulate the expected spectrum at the near and far detectors, using the unoscillated spectrum from Fig. 3 of \cite{Adey:2018zwh}.  As the energy smearing was already included, we performed a Gaussian convolution only for the oscillation probability, following the procedure described in \cite{Ciuffoli:2019nli}.  As explained there, the errors due to this this approximation are minimal, since it is used both in the Asimov data set and in the fit.

The aim of this section is to show that the decoherence, if ignored, could affect the measurement of $\theta_{13}$. For this reason we included decoherence in the generation of the Asimov data set but assumed no decoherence when we fit the data.  In the fit, we used the two-flavor oscillations probability 
\[
P_{ee}=1-\textrm{sin}^2(2\theta_{13})\textrm{sin}^2(1.27\Delta m_{ee}^2 L/E).
\]
The oscillation probability used for the Asimov data set was modified to include
\[
P_{ee;d}=1-e^{-L^2/(2D^2)}\textrm{sin}^2(2\theta_{13})\textrm{sin}^2(1.27\Delta m_{ee}^2 L/E)
\]
where $D$ is the decoherence length. $\Delta m_{ee}$ was held fixed and equal to $2.482\times 10^{-3}\textrm{eV}^2$ both in the Asimov data set and in the fit, while the value of $\theta_{13}$ used for the computation of the Asimov data set was $\textrm{sin}^2(2\theta_{13})=0.085$. An additional pull parameter was introduced to take into account for the uncertanity in the total flux normalization.  No penalty terms were considered.

\begin{figure} 
\begin{center}
\includegraphics[width=2.5in,height=1.7in]{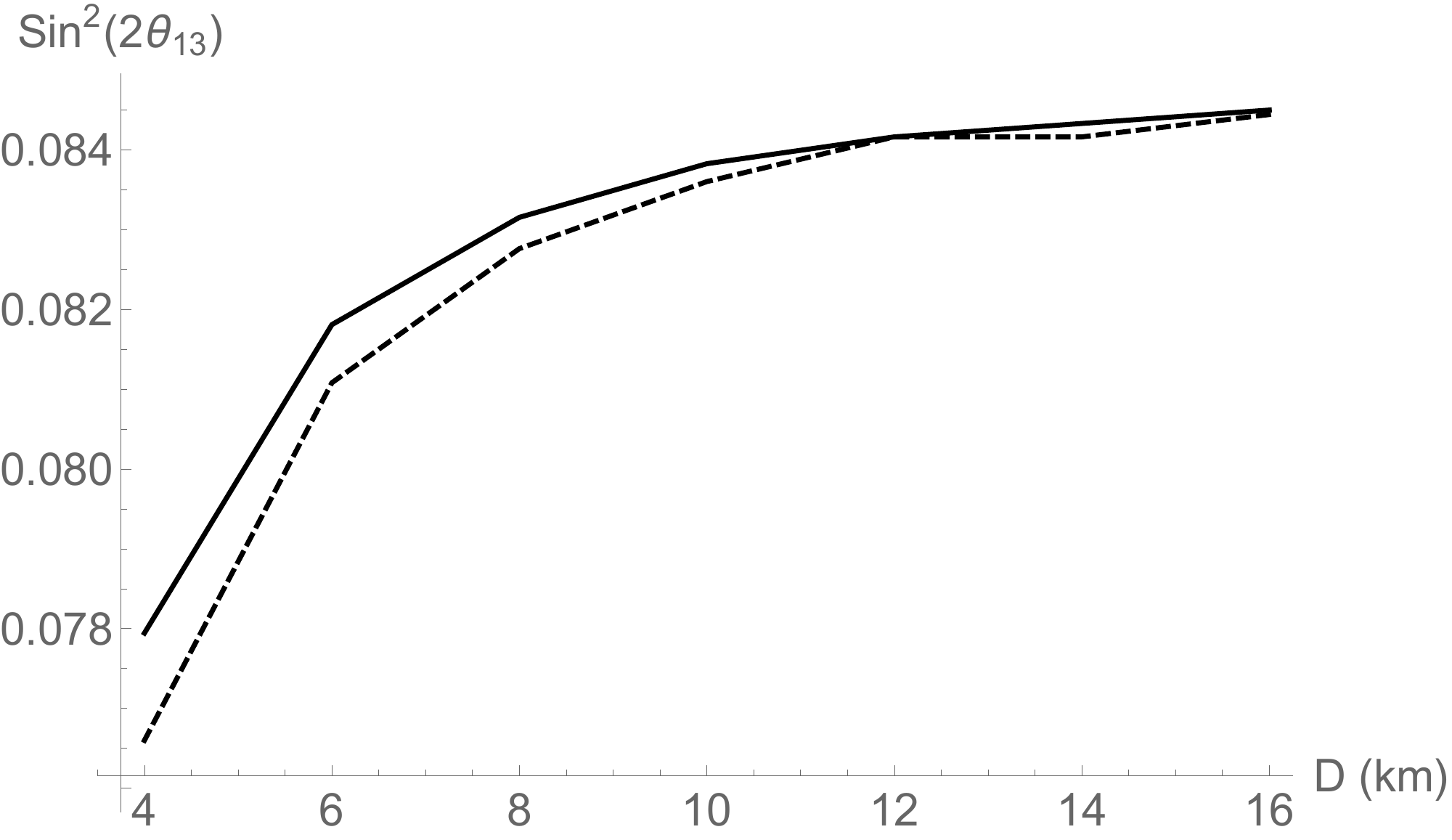}
\includegraphics[width=2.5in,height=1.7in]{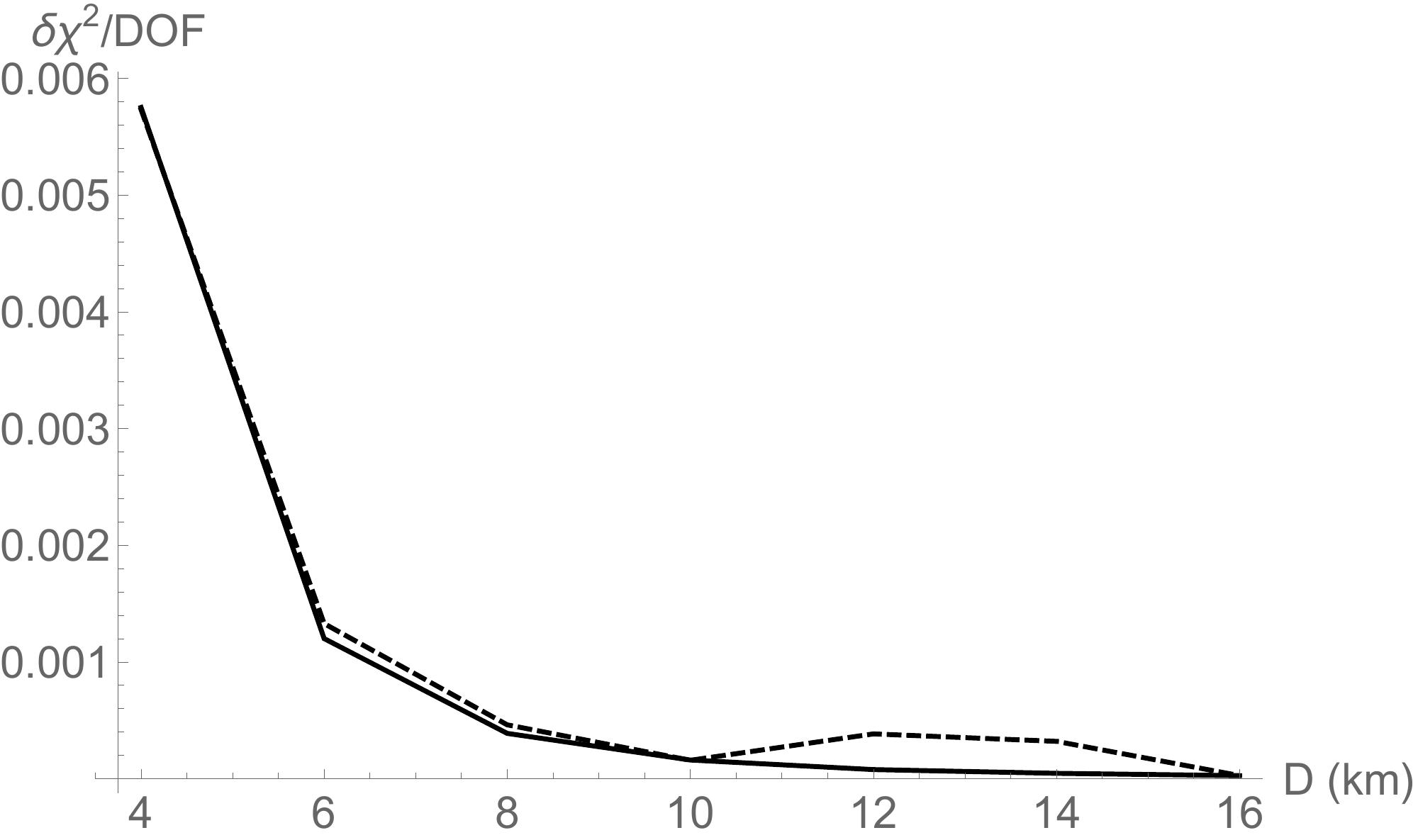}
\caption{Left Panel: Best-Fit value for $\textrm{sin}^2(2\theta_{13})$, as a function of the decoherence length. Right panel: expected increment of $\chi^2/DOF$ if the decoherence is not taken into account. Dashed curves: rate-only analysis. Solid curves: rate+shape analysis}
\label{dech}
\end{center}
\end{figure}

We used both a rate-only and rate+shape analysis (the use of a pull parameter for the total normalization does not invalidate the rate only analysis, due to the presence of near detectors). In Fig. \ref{dech} (left panel) it is possible to see that the best-fit value for $\textrm{sin}^2(2\theta_{13})$ is significantly different for finite values of $D$. For a comparison, it should be underlined that in the analysis performed in \cite{An:2016ses}, the 1-$\sigma$ range found for $\textrm{sin}^2(2\theta_{13})$ (taking into account only the statistical fluctuations and not the systematical errors) was 0.0030 for the rate-only analysis and 0.0027 for the rate+shape analysis. 

If the model used to fit the experimental data is correct, asymptotically the $\chi^2/DOF$ should approach to 1, however if some effects are not taken into account, this could increase the value of the $\chi^2 $, leading to a higher $\chi^2/DOF$ ratio. Since we used two different models for the Asimov data set and the fit, the minimum of the $\chi^2$ is not zero: this number represents the expected incrementation of the $\chi^2$ statistic (after minimizing over $\theta_{13}$ and all the other pull parameters present) for not considering decoherence in the fit. We will call min$\chi^2=\delta\chi^2$; asymptotically $\chi^2/DOF$ will now approach $1+\delta\chi^2/DOF$; if the deviation from 1 is significant, this means that a fit would be noticeably worse and it would be possible to realize that the model used to fit the data is not correct. In Fig. \ref{dech} (right panel) we report $\delta \chi^2/DOF$: from the plot we can notice that this quantity is quite small, and this deviation could go unnoticed in a goodness-of-fit test. In the rate-only analysis, the number of degrees of freedom is 6: 8 data points, one from each detector, minus 2, which is the number of parameters minimized, namely the total flux normalization and $\theta_{13}$, while in the rate+shape analysis we divided the spectrum between 1.8 and 7.8 MeV into 24 0.25-MeV energy bins, increasing the number of degrees of freedom up to 190 ($=24\times 8-2$).


\begin{thebibliography}{99}

\bibitem{naumov1}
  D.~V.~Naumov and V.~A.~Naumov,
  ``A Diagrammatic treatment of neutrino oscillations,''
  J.\ Phys.\ G {\bf 37} (2010) 105014
  doi:10.1088/0954-3899/37/10/105014
  [arXiv:1008.0306 [hep-ph]].

\bibitem{naumov2}
  D.~V.~Naumov,
  ``On the Theory of Wave Packets,''
  Phys.\ Part.\ Nucl.\ Lett.\  {\bf 10} (2013) 642
  doi:10.1134/S1547477113070145
  [arXiv:1309.1717 [quant-ph]].

\bibitem{dayadec}
  F.~P.~An {\it et al.} [Daya Bay Collaboration],
  ``Study of the wave packet treatment of neutrino oscillation at Daya Bay,''
  Eur.\ Phys.\ J.\ C {\bf 77} (2017) no.9,  606
  doi:10.1140/epjc/s10052-017-4970-y
  [arXiv:1608.01661 [hep-ex]].

\bibitem{t2k}
  K.~Abe {\it et al.} [T2K Collaboration],
  ``Search for CP Violation in Neutrino and Antineutrino Oscillations by the T2K Experiment with $2.2\times10^{21}$ Protons on Target,''
  Phys.\ Rev.\ Lett.\  {\bf 121} (2018) no.17,  171802
  doi:10.1103/PhysRevLett.121.171802
  [arXiv:1807.07891 [hep-ex]].


\bibitem{steven}
  Y.~L.~Chan, M.-C.~Chu, K.~M.~Tsui, C.~F.~Wong and J.~Xu,
  ``Wave-packet treatment of reactor neutrino oscillation experiments and its implications on determining the neutrino mass hierarchy,''
  Eur.\ Phys.\ J.\ C {\bf 76} (2016) no.6,  310
  doi:10.1140/epjc/s10052-016-4143-4
  [arXiv:1507.06421 [hep-ph]].



\bibitem{beuthe}
  M.~Beuthe,
  ``Oscillations of neutrinos and mesons in quantum field theory,''
  Phys.\ Rept.\  {\bf 375} (2003) 105
  doi:10.1016/S0370-1573(02)00538-0
  [hep-ph/0109119].

\bibitem{giunti2012}
 C.~Giunti,
  ``Neutrino wave packets in quantum field theory,''
  JHEP {\bf 0211} (2002) 017
  doi:10.1088/1126-6708/2002/11/017
  [hep-ph/0205014].

\bibitem{noi1}
  J.~Evslin, H.~Mohammed, E.~Ciuffoli and Y.~Zhou,
  ``Entangled Neutrino States in a Toy Model QFT,''
  Eur.  Phys. J. C In Press,
  arXiv:1902.03934 [hep-ph].

\bibitem{nussinov76}
  S.~Nussinov,
  ``Solar Neutrinos and Neutrino Mixing,''
  Phys.\ Lett.\  {\bf 63B} (1976) 201.
  doi:10.1016/0370-2693(76)90648-1

 \bibitem{daya18}
 D.~Adey {\it et al.} [Daya Bay Collaboration],
  ``Measurement of the Electron Antineutrino Oscillation with 1958 Days of Operation at Daya Bay,''
  Phys.\ Rev.\ Lett.\  {\bf 121} (2018) no.24,  241805
  doi:10.1103/PhysRevLett.121.241805
  [arXiv:1809.02261 [hep-ex]].

\bibitem{t2k}
  H.~O'Keeffe [T2K Collaboration],
  ``Recent T2K Neutrino Oscillation Results,''
  PoS LeptonPhoton {\bf 2019} (2019) 098.
  doi:10.22323/1.367.0098

\bibitem{An:2016ses} 
  F.~P.~An {\it et al.} [Daya Bay Collaboration],
  ``Measurement of electron antineutrino oscillation based on 1230 days of operation of the Daya Bay experiment,''
  Phys.\ Rev.\ D {\bf 95}, no. 7, 072006 (2017)
  doi:10.1103/PhysRevD.95.072006
  [arXiv:1610.04802 [hep-ex]].
\bibitem{Adey:2018zwh} 
  D.~Adey {\it et al.} [Daya Bay Collaboration],
  ``Measurement of the Electron Antineutrino Oscillation with 1958 Days of Operation at Daya Bay,''
  Phys.\ Rev.\ Lett.\  {\bf 121}, no. 24, 241805 (2018)
  doi:10.1103/PhysRevLett.121.241805
  [arXiv:1809.02261 [hep-ex]].
\bibitem{Ciuffoli:2019nli} 
  E.~Ciuffoli, J.~Evslin and H.~Mohammed,
  ``Uncertainty in the Reactor Neutrino Spectrum and Mass Hierarchy Determination,''
  JHEP {\bf 1910}, 143 (2019)
  doi:10.1007/JHEP10(2019)143
  [arXiv:1907.02309 [hep-ph]].

\end{thebibliography}

\begin{thebibliography}{99}

\bibitem{boya2011}
  D.~Boyanovsky,
  ``Short baseline neutrino oscillations: when entanglement suppresses coherence,''
  Phys.\ Rev.\ D {\bf 84} (2011) 065001
  doi:10.1103/PhysRevD.84.065001
  [arXiv:1106.6248 [hep-ph]].


\bibitem{accdec}
  B.~J.~P.~Jones,
  ``Dynamical pion collapse and the coherence of conventional neutrino beams,''
  Phys.\ Rev.\ D {\bf 91} (2015) no.5,  053002
  doi:10.1103/PhysRevD.91.053002
  [arXiv:1412.2264 [hep-ph]].

\bibitem{steven}
  Y.~L.~Chan, M.-C.~Chu, K.~M.~Tsui, C.~F.~Wong and J.~Xu,
  ``Wave-packet treatment of reactor neutrino oscillation experiments and its implications on determining the neutrino mass hierarchy,''
  Eur.\ Phys.\ J.\ C {\bf 76} (2016) no.6,  310
  doi:10.1140/epjc/s10052-016-4143-4
  [arXiv:1507.06421 [hep-ph]].


\bibitem{dayadec}
  F.~P.~An {\it et al.} [Daya Bay Collaboration],
  ``Study of the wave packet treatment of neutrino oscillation at Daya Bay,''
  Eur.\ Phys.\ J.\ C {\bf 77} (2017) no.9,  606
  doi:10.1140/epjc/s10052-017-4970-y
  [arXiv:1608.01661 [hep-ex]].


\bibitem{revival}
  K.~Kiers and N.~Weiss,
  ``Neutrino oscillations in a model with a source and detector,''
  Phys.\ Rev.\ D {\bf 57} (1998) 3091
  doi:10.1103/PhysRevD.57.3091
  [hep-ph/9710289].

\bibitem{nuss76}
  S.~Nussinov,
  ``Solar Neutrinos and Neutrino Mixing,''
  Phys.\ Lett.\  {\bf 63B} (1976) 201.
  doi:10.1016/0370-2693(76)90648-1

\bibitem{wilczek}
  L.~Krauss and F.~Wilczek,
  ``Solar Neutrino Oscillations,''
  Phys.\ Rev.\ Lett.\  {\bf 55} (1985) 122.
  doi:10.1103/PhysRevLett.55.122

\bibitem{rich}
 J.~Rich,
  ``The Quantum mechanics of neutrino oscillations,''
  Phys.\ Rev.\ D {\bf 48} (1993) 4318.
  doi:10.1103/PhysRevD.48.4318

\bibitem{boriserr}
  B.~Kayser and J.~Kopp,
  ``Testing the wave packet approach to neutrino oscillations in future experiments,''
  arXiv:1005.4081 [hep-ph].

\bibitem{giunti2012}
 C.~Giunti,
  ``Neutrino wave packets in quantum field theory,''
  JHEP {\bf 0211} (2002) 017
  doi:10.1088/1126-6708/2002/11/017
  [hep-ph/0205014].

\bibitem{zurek}
  W.~H.~Zurek,
  ``Environment induced superselection rules,''
  Phys.\ Rev.\ D {\bf 26} (1982) 1862.
  doi:10.1103/PhysRevD.26.1862

\bibitem{mcgreevy}
A.~Kobach, A.~V.~Manohar and J.~McGreevy,
  ``Neutrino Oscillation Measurements Computed in Quantum Field Theory,''
  Phys.\ Lett.\ B {\bf 783} (2018) 59
  doi:10.1016/j.physletb.2018.06.021
  [arXiv:1711.07491 [hep-ph]].

\bibitem{cgl}
  A.~G.~Cohen, S.~L.~Glashow and Z.~Ligeti,
  ``Disentangling Neutrino Oscillations,''
  Phys.\ Lett.\ B {\bf 678} (2009) 191
  doi:10.1016/j.physletb.2009.06.020
  [arXiv:0810.4602 [hep-ph]].

\bibitem{mcdonald}
``Oscillations and decoherence,"
Kirk T McDonald, Talk at NuFact 2013,
August 23, 2013, Beijing, China.

\bibitem{naumov1}
  D.~V.~Naumov and V.~A.~Naumov,
  ``A Diagrammatic treatment of neutrino oscillations,''
  J.\ Phys.\ G {\bf 37} (2010) 105014
  doi:10.1088/0954-3899/37/10/105014
  [arXiv:1008.0306 [hep-ph]].


\bibitem{naumov2}
  D.~V.~Naumov,
  ``On the Theory of Wave Packets,''
  Phys.\ Part.\ Nucl.\ Lett.\  {\bf 10} (2013) 642
  doi:10.1134/S1547477113070145
  [arXiv:1309.1717 [quant-ph]].


\bibitem{akqft}
  E.~K.~Akhmedov and A.~Y.~Smirnov,
  ``Neutrino oscillations: Entanglement, energy-momentum conservation and QFT,''
  Found.\ Phys.\  {\bf 41} (2011) 1279
  doi:10.1007/s10701-011-9545-4
  [arXiv:1008.2077 [hep-ph]].

\bibitem{akmoss}
  E.~K.~Akhmedov, J.~Kopp and M.~Lindner,
  ``Oscillations of Mossbauer neutrinos,''
  JHEP {\bf 0805} (2008) 005
  doi:10.1088/1126-6708/2008/05/005
  [arXiv:0802.2513 [hep-ph]].


\bibitem{colemanlect}
 B.~G.~G.~Chen, D.~Derbes, D.~Griffiths, B.~Hill, R.~Sohn and Y.~S.~Ting,
  ``Lectures of Sidney Coleman on Quantum Field Theory,''
  doi:10.1142/9371 .


\bibitem{review}
  M.~Beuthe,
  ``Oscillations of neutrinos and mesons in quantum field theory,''
  Phys.\ Rept.\  {\bf 375} (2003) 105
  doi:10.1016/S0370-1573(02)00538-0
  [hep-ph/0109119].



\end{thebibliography}
\end{document}